\documentclass[useAMS,usenatbib]{mn2e}
\usepackage{graphicx}
\usepackage{eufrak}
\usepackage{eucal}
\usepackage{txfonts}
\usepackage{float}

\newfloat{code}{h}{lop}

\begin{document}

\title[On the dust abundance gradients in late-type galaxies II]{On the dust abundance gradients in late-type galaxies:\\
II. Analytical models as evidence for massive interstellar dust growth in SINGS galaxies}

\author[Mattsson \& Andersen]{Lars Mattsson\thanks{E-mail: mattsson@dark-cosmology.dk} \& Anja C. Andersen\\
Dark Cosmology Centre, Niels Bohr Institute, University of Copenhagen, Juliane Maries Vej 30, DK-2100, Copenhagen \O, Denmark}

\date{}

\pagerange{\pageref{firstpage}--\pageref{lastpage}} \pubyear{2011}

\maketitle

\label{firstpage}

\begin{abstract}
We use simple analytical models of the build up of the dust component and compare these with radial dust distributions derived from observations
of SINGS galaxies. The observations show that dust gradients are indeed typically steeper than the corresponding metallicity gradients and our models indicate very little dust 
destruction, but significant dust growth in the ISM for most of these galaxies. Hence, we conclude that there is evidence for significant non-stellar dust production, and little evidence 
for dust destruction due to SNe shock waves. We find that dust is reprocessed rather than destroyed by shocks from SNe. Finally, we argue that dust abundances
derived using standard methods may be overestimated, since even very 'generous' estimates of the metallicity {results in}  dust-to-metals ratios above unity in several cases, {if the dust abundances given in the literature are taken at face value.}
\end{abstract}


\section{Introduction}
Dust grains in dense molecular can theoretically grow to micrometer sizes by accretion and coagulation \citep[see, e.g.,][]{Ossenkopf93,Ormel09,Hirashita11}.
There is also observational evidence suggesting such large grains are abundant in many Galactic molecular clouds \citep[see][and references therein]{Pagani10},
which is easily explained by efficient dust growth in these cores.
Micrometer-sized dust grains can theoretically be formed in carbon rich AGB stars as well \citep{Mattsson11b}, and are likely to form also in oxygen rich AGB stars
\citep{Hoefner08}, but the probability of having any larger quantities of these grains gathering in dense molecular cores is very low. Furthermore, grains are expected 
to survive in the Galactic ISM for a few hundred Myr \citep{Jones96,Jones04,Serra08,Jones11}, while the time scale for any significant stellar dust enrichment
in the local ISM is about Gyr, which implies the presence of dust growth processes in the ISM. Hence, interstellar dust growth can be regarded an established phenomenon.

Shock-waves from supernovae (SNe) is thought to be able to {destroy} dust grains as these waves propagate through 
the interstellar medium (ISM). The time scale for this dust destruction is {largely dependent on} the supernova rate (SNR) and how much dust 
each SN-shock can destroy \citep{McKee89, Draine90}. This shock destruction of dust grains {has been} considered quite efficient 
\citep{Jones96,Jones04}, {although more recent work show that while this is likely the case for carbon dust it may not be the case for silicates 
\citep{,Serra08,Jones11}} which seems consistent with the Milky Way (solar neighbourhood). Very efficient dust destruction {seems on the other
hand} inconsistent with the high dust masses detected in high-$z$ objects \citep{Dwek07, Gall11, Mattsson11}. It may of course be that dust 
destruction by SNe is less efficient in high-$z$ galaxies, but also modelling of nearby late-type galaxies seems to work nicely without significant net 
destruction of dust \citep{Inoue03, Hirashita99}, and it makes it easier to explain the dust-to-gas ratios, since stellar dust 
production is insufficient \citep{Hirashita99, Zhukovska08}. 

Observations imply that dust production in SNe is rather inefficient \citep{Kotak06,Kotak09}, which suggest the large masses of cold dust
detected in some SN remnants \citep[see, e.g.,][]{Morgan03,Morgan03b,Dunne09} may be the result of subsequent dust growth rather than 
an effect of heating or actual dust production in the actual SN. Theoretical results suggest 90\% of the dust produced in SNe is destroyed by the 
reverse shock before it reaches the ISM \citep{Bianchi07}, which leads to a relatively consistent picture where AGB stars is the most important source 
of stellar dust \citep{Edmunds98,Ferrarotti06}. Models of the dust evolution in the solar neighbourhood by \citet{Dwek98} and \citet{Zhukovska08} 
suggest dust production in stars and possible dust destruction by SN-shock waves need to be compensated by an efficient dust growth in the ISM. 
This is also supported by the fact that dust grains of considerable sizes seem to exist in the ISM \citep{Pagani10}. However, recent results from 
Herchel observations of SN 1987A show a huge reservoir of cold dust which suggest that at least some SNe may produce
significant amounts of dust \citep{Matsuura11}. However, it is still unclear how much of this dust will survive once the reverse shock sets in.

It is obviously not clear whether dust destruction from SN-shock waves in the ISM is an efficient mechanism or not, or perhaps more accurately put: whether the {\it net}
dust destruction is important or not. As mentioned above, existing models of the evolution of the dust phase in the solar neighbourhood (solar circle) rely on the assumption 
that both growth and destruction are important \citep{Dwek98,Zhukovska08}. But with only a single data point, e.g., the present-day dust abundance in the solar neighbourhood, 
it is hard to tell  whether a model with, or without, dust destruction by SNe and/or dust growth in the ISM makes a better fit to data. Recently, however, results from the SINGS 
project \citep{Kennicutt03} have provided dust  abundance profiles based on physical dust models for a good number of nearby galaxies 
\citep{Munoz-Mateos09} and the associated THINGS project has provided H~\textsc{i} profiles  for many of the SINGS galaxies \citep{Walter08}. Hence, 
the dust-to-gas ratio along the disc can be derived for many of these galaxies, which opens for more precise comparisons with observations since 
different galactocentric distances in the discs represents different evolutionary states of the interstellar gas. Many of the late-type galaxies in this 
sample have also well-constrained metallicity gradients as there are sufficient numbers of detectable H\textsc{ii}-regions for which spectroscopy 
have been done \citep{Moustakas10}. 

Previous work on dust evolution modelling has shown that the dust-to-metals ratio may change as the evolution proceeds. The analytical
relations derived by, e.g., \citet{Edmunds01} and \citet{Mattsson11}, show that the dust abundance may not necessarily follow the metal
abundances in the ISM. More detailed results, obtained through numerical modelling, suggest the same \citep{Dwek98,Inoue03,Zhukovska08}.
The reasons for the changing dust-to-metals ratio are mainly the destruction (including astration) and growth of dust grains that may occur in the ISM,
which suggest that dust and metallicity gradients may not have the same slope, i.e., there is a {\it dust-to-metals gradient}. As suggested in 
\citet[][henceforth 'Paper I']{Mattsson11c} the dust-to-metals gradient along a galactic disc can essentially be regarded as a diagnostic for net dust growth or
net destruction depending on whether the slope is negative or positive, respectively. 

We demonstrate here a way to estimate the efficiency of the net dust 
destruction by SNe in nearby late-type galaxies using dust- and metallicity gradients.  Comparing the dust abundance gradients along the disc 
with the metallicity gradients may give an indication of whether net destruction or net growth of dust is important or not, as we will show in this paper.

\section{Model}
\label{theory}
\subsection{Basic equations}
Following Paper I we use the instantaneous recycling approximation \cite[IRA, which essentially means all stars are assumed to have negligible 
lifetimes {compared to the time scale for the build-up of metals and dust}, see][]{Pagel97} throughout this paper. 
Then, assuming a 'closed box', gas and dust destruction in the ISM from SN-shocks, the equations for the metallicity $Z$ and the dust-to-gas ratio $Z_{\rm d}$ becomes
\begin{equation}
\label{metals}
\Sigma_{\rm g}{d Z \over d t} = y_{Z}{d\Sigma_{\rm s}\over dt} = -y_{Z}{d\Sigma_{\rm g}\over dt},
\end{equation}
\begin{equation}
\label{dust}
\Sigma_{\rm g}{d Z_{\rm d} \over d t} = y_{\rm d}{d\Sigma_{\rm s}\over dt} + Z_{\rm d}(r,t)\left({1\over \tau_{\rm gr}} - {1\over \tau_{\rm d}}\right),
\end{equation}
where $\tau_{\rm d}$ is the dust destruction time scale,  $\tau_{\rm gr}$ is the dust-growth time scale and $y_{\rm d}$, $y_Z$ denote the yields (dust and metals, respectively) as defined in Paper I, {and $\Sigma_{\rm s}$, $\Sigma_{\rm g}$ denotes surface
density by mass of stars and gas, respectively.} Combining equations (\ref{metals}) and (\ref{dust}), we have
\begin{equation}
{\partial Z_{\rm d}\over \partial Z} = {y_{\rm d}+Z_{\rm d}(\tau^{-1}_{\rm gr}-\tau^{-1}_{\rm d}) \over y_Z},
\end{equation}
which thus have no explicit dependence on the gas mass density $\Sigma_{\rm g}$ or the stellar mass density $\Sigma_{\rm s}$, {although $Z$,
$Z_{\rm d}$ as well as the time scales $\tau_{\rm gr}$, $\tau_{\rm d}$ may {\it implicitly} depend on $\Sigma_{\rm g}$ and $\Sigma_{\rm s}$.}.

\subsection{Destruction and growth of dust  in the ISM}
\label{dustdestgrowth}
Following \citet{McKee89} the dust destruction time-scale is
\begin{equation}
\tau_{\rm d} = {\Sigma_{\rm g}\over \langle m_{\rm ISM}\rangle\,R_{\rm SN}},
\end{equation}
where $\langle m_{\rm ISM}\rangle$ is the effective gas mass cleared of dust by each SN event {(which is not necessarily the same as the
gas mass affected by each SN)}, and $R_{\rm SN}$
is the SN rate. As shown in Paper I, the time scale $\tau_{\rm d}$ may be expressed as
\begin{equation}
\label{taud}
\tau_{\rm d}^{-1} \approx  {\delta\over \Sigma_{\rm g}}{d\Sigma_{\rm s}\over dt},
\end{equation}
where $\delta$ will be referred to as the {dust destruction parameter, which is a kind of measure of the efficiency of dust destruction.} For a 
\citet{Larson98} IMF and $m_{\rm ISM} \approx 1000 M_\odot$ \citep{Jones96,Jones04}, then $\delta \approx 10$ \citep[see][]{Mattsson11}. 

We define the rate per unit volume at which the number of atoms in dust grains may
{\it grow} by accretion of metals onto these dust grains as \citep[see, e.g.][]{Dwek98} at a rate which naturally depends on metallicity $Z$, the typical grain radius $a$ and the sticking
coefficient $f_{\rm s}$ (i.e., the probability that an atom will stick to the grain).  Additionally, the mean thermal speed of the gas particles (including metals) $\langle v_{\rm g}\rangle$
will affect this rate, which suggest a temperature dependence. The timescale of grain growth can thus be expressed as (again, see Paper I for brief derivation)
\begin{equation}
\label{taugr}
\tau_{\rm gr} = \tau_0 \left(1- {Z_{\rm d}\over Z}\right)^{-1}, \quad \tau_0 
                       = {\langle m_{\rm gr}\rangle\,d_{\rm c}\over f_{\rm s}\pi a^2 Z\,\Sigma_{\rm mol} \langle v_{\rm g}\rangle},
\end{equation}
{where $\langle m_{\rm gr}\rangle$ is the average mass of individual dust grains in the gas, $d_{\rm c}$ is the characteristic size of the molecular
clouds where dust grows and $\Sigma_{\rm mol} $ is the surface density of molecular gas}. As in Paper I, we
 will assume $\Sigma_{\rm mol} \approx \Sigma_{\rm H_2} $, since most of the gas in molecular gas
clouds is in the form of molecular hydrogen. We also assume $\Sigma_{\rm H_2} $ traces the star-formation rate
\citep{Rownd99,Wong02,Bigiel08,Leroy08,Bigiel11,Feldmann11,Schruba11}. This is also supported by theory and recent numerical 
experiments \citep[see, e.g.][]{Krumholz09,Krumholz11}. Moreover, the mean thermal speed $\langle v_{\rm g}\rangle$ is roughly constant in the 
considered ISM environment 
and the typical grain radius does not vary much. Hence, as shown in Paper I, $\tau_0$ is essentially just a simple function of the metallicity and the growth rate of the stellar component,
\begin{equation}
\tau_0 ^{-1}= {\epsilon Z\over \Sigma_{\rm g}} {d\Sigma_{\rm s}\over dt},
\end{equation}
where $\epsilon$ will, in the following, be treated as a free {(but not unconstrained - see Paper I)} parameter of the model. The rate of
change of the dust-to-gas ratio $Z_{\rm d}$ due to accretion of metals onto pre-existing dust grains in the ISM is then
\begin{equation}
\left({dZ_{\rm d}\over dt}\right)_{\rm gr}  = \left(1-{Z_{\rm d}\over Z}\right) {Z_{\rm d}\over \tau_{\rm gr}}.
\end{equation}
{The $\epsilon$ used above should also not be confused with the dust-desruction efficiency 
parameter used by \citet{McKee89} and it also differs slightly from the dust-growth parameter used in \citet{Mattsson11} as we in this work consider
the free atomic metals rather than the total metallicity.}

\subsection{Models of radial dust-to-gas distributions}
\label{models}
For simplicity we stick to a closed-box model as in Paper I, i.e., no in- or outflows to/from the disc will be considered. This is not in agreement with the widely accepted ideas about 
galaxy-disc formation, where the baryons (in
the form of essentially pristine gas) are assumed to be accreted over an extended period of time. But as shown by \citet{Edmunds90}, the most prominent effect of unenriched infall is to make the
effective yield smaller, i.e., to dilute the gas so that the metallicity builds up more slowly. As we use the present-day metallicity as input (see section \ref{modelresults}), the overall effects
of assuming a closed box are rather small. Note also, that \citet{Garnett97} have showed that the simple closed-box model of chemical evolution seem to work quite well for modelling 
oxygen (O/H) gradients. In particular, adopting the \citet{Clayton87} infall model neither means a significant improvement of the fit, nor a very large change of the required yield.

Adopting a closed-box scenario, the dust destruction and dust growth models as described above, we arrive at the equation
\begin{equation}
\label{dustz}
{dZ_{\rm d}\over dZ} = {1\over y_Z}\left\{y_{\rm d} + Z_{\rm d} \left[\epsilon \left(1-{Z_{\rm d}\over Z} \right)\,Z -\delta\right] \right\},
\end{equation}
where $y_Z$ is the metal yield. With $0\le y_{\rm d}\le y_Z$ as a basic requirement, the general closed-box solution (of equation \ref{dustz}) for the dust-to-gas ratio $Z_{\rm d}$ in terms of the metallicity $Z$ can then
be expressed in terms of the confluent hypergeometric Kummer-Tricomi functions of the first and second kind (usually denoted $U$ and $M$), respectively \citep{Kummer1837,Tricomi47}. 
We refer to Paper I for further details. In case there is no dust destruction by SNe ($\delta = 0$) the solution reduces to
\begin{equation}
\label{growthsol}
Z_{\rm d} = {y_{\rm d}\over y_Z} {
M\left(1+{1\over 2}{y_{\rm d}\over y_Z}, {3\over 2}; {1\over 2}{\epsilon Z^2\over y_Z}\right)
\over 
M\left({1\over 2}{y_{\rm d}\over y_Z}, {1\over 2}; {1\over 2}{\epsilon Z^2 \over y_Z}\right)}
\,Z,
\end{equation}
where $M(a,b;z)={}_1F_1(a,b;z)$ is the Kummer-Tricomi function of the first kind, which is identical to the confluent hypergeometric function $_1F_1(a,b;z)$ that is often implemented in computer algebra
software. If there is neither growth, nor destruction of dust in the ISM ($\epsilon = \delta = 0$), we have the trivial case
\begin{equation}
\label{stellarsol}
Z_{\rm d} = {y_{\rm d} \over y_Z} Z,
\end{equation}
corresponding to pure stellar dust production. These are the two cases we consider in detail in this paper.

\section{Results and discussion}
We have compared the theory of interstellar dust evolution presented in Paper I with data on dust gradients in late-type galaxies taken from the analysis of SINGS sample by \citet{Munoz-Mateos09}.
They present radial distributions of dust (surface densities) in galactic discs derived from UV-, IR- and gas mass profiles, which have relatively small errors. These data are likely the best constraints on
radial dust gradients in galactic discs available at present. Here we summarise and discuss the results.

\subsection{General results and limitations}
In a previous paper (Paper I) we have shown that dust destruction by shocks from exploding SNe will flatten a negative (declining with galactocentric distance) dust-to-gas gradient over a galactic disc, 
while dust growth in general acts as to make it steeper (see the Theorem proved in Paper I). From this result, we expect dust-to-metals gradients to have {\it positive} gradients (rising with galactocentric 
distance) if dust destruction is more important than dust growth, and if dust growth is the more important process we expect them to be {\it negative}.

In our limited sample of late-type galaxies taken from \citet{Munoz-Mateos09}, we have found that the dust-to-metals gradients are typically negative  or in some cases flat. 
There is only one counter-example (NGC 5194) where the metallicity gradient may be somewhat steeper than the dust-to-gas gradient (see figure \ref{grad}), resulting in a positive dust-to-metals 
gradient (see figure \ref{zeta}). However, it should be stressed that the metallicity gradients (as traced by the O/H gradients in the discs) we adopt in this study may not be correct in all cases. We
have taken the metallicity gradients from \citet{Moustakas10}, which are derived using the strong-line calibration by \citet{Kobulnicky04}, except for the dwarf irregular galaxy Holmberg II. For the
latter we assume an essentially flat metallicity gradient with a slope of only 1\%, since dwarf galaxies typically have no metallicity gradients, but the slope cannot be exactly zero if a dust-to-gas/metals
gradient is to be obtained from the model. \citet{Moustakas10} also provide gradients based on 
abundances derived using the $P$-method \citep{Pilyugin05}, which suggest significantly lower metallicities and somewhat flatter gradients. Using the latter, the amount of metals are clearly 
inconsistent with the dust abundances derived by \citet{Munoz-Mateos09}, i.e., dust-to-metals ratios are above unity for a majority of data points. Therefore, we present only models (see section \ref{modelresults}) 
based on metallicity gradients derived using the \citet{Kobulnicky04} calibration, which are also closer to the dust-to-gas gradients in terms of steepness (cf. the models with pure stellar dust production in Figure
\ref{grad}, which in several cases are relatively close to the dust-to-gas gradients). 

This inconsistency (that the dust-to-metals ratio exceeds unity), is indeed a problem, and the reader should bear in mind that we 
have not used these metallicity data because they are the most reliable, but rather because they are not obviously inconsistent with the dust-to-gas ratios we compare with. As will be shown below, 
there are still a few data points at small galactocentric distances which would correspond to dust-to-metals ratios above unity also when using the metallicity gradients derived using the \citet{Kobulnicky04} calibration.  
Either the dust abundances should  be about a factor of 2-3 lower, or the metal abundance should be increased by a similar factor. The latter is less likely, however, since we already use oxygen abundances which are 
at the high end of the possible range (set by the various methods and calibrations for abundance determination) to estimate the metallicity along the disc and assume a relatively low oxygen fraction. Hence, it is 
more likely the dust abundance is overestimated in general, which is an interpretation that is supported by recent results on the dust content of dwarf galaxies \citep{Madden11} and which we will discuss in more 
detail in section \ref{dustestimates}. But for now, in order to avoid constructing strange inconsistent models, we have just scaled down the dust abundances for some galaxies such that the dust-to-metals ratio never
exceeds 0.9 anywhere along the discs. 

\subsection{Models and fitting}
\label{modelresults}
We have fitted simple analytical models (see section \ref{models}) to the dust-to-gas profiles derived from observations by \citet{Munoz-Mateos09}. In doing so we have used the Levenberg-Markwardt scheme
for $\chi^2$-minimisation. More precisely, we have used the IDL-routine package MPFIT \citep[][]{Markwardt09} in combination with a numerical implementation (for IDL) of the Kummer-Tricomi functions
(see Paper I). This setup has proven very stable for the cases we consider here (equations \ref{growthsol} and \ref{stellarsol}).

\subsubsection{Input data}
For the model fits we have fixed the metal yield $y_Z$ to the value obtained from a simple closed-box scenario
\begin{equation}
y_Z = {Z(R=0.4R_{25})\over \ln(1/\mu)},
\end{equation}
where $\mu$ is the overall gas mass fraction of the galaxy (see Table \ref{parameters}). The metallicity at a galactocentric distance $R=0.4R_{25}$ is known to be a good proxy for the typical metallicity of a galaxy
disc \citep{Garnett02}, which is why we define the effective metal yield as above. We have used the same gas mass fractions as in \citet{Pilyugin04}.
Note that the stellar dust yield $y_{\rm d}$, on the other hand, is treated as a free parameter.

Furthermore, we use oxygen (O/H) gradients derived using the strong-line calibration by \citet{Kobulnicky04} as input to our model \citep[see][for further details]{Moustakas10}. 
We use the log-linear-fits and do not consider possible bends and features of the metallicity profiles as metallicities derived from H~\textsc{ii} regions can be quite uncertain.
To obtain the total metal fraction (metallicity) we first convert the number abundances of oxygen into oxygen mass fractions using the relation \citep{Garnett02}
\begin{equation}
X_{\rm O} = 12\, {\rm O\over H} ,
\end{equation}
in which we then have implicitly assumed $M_{\rm gas} = 1.33 M_{\rm H}$. Then, we assume that oxygen typically makes up about a third of all metals \citep[this is certainly at the low end of the possible 
range, see e.g.,][where 45-60\% is the suggested value]{Garnett02}, hence,
\begin{equation}
Z = 36\, {\rm O\over H} .
\end{equation}
The fraction of oxygen in the metals is of course not a universal constant and may vary from one environment to another. In particular, the oxygen fraction is likely higher in low-metallicty
environments than it is in high-metallicity environments where low- and intermediate stars as well as SNe type Ia have contributed significantly to the metal enrichment of the ISM.

\subsubsection{Modelling results}
Allowing dust destruction ($\delta \ge 0$) does not seem to improve the agreement between model and the dust-to-gas profile inferred from observations. For example, we considered the case where
$\delta$ is fixed to a specific value (we assumed $\delta = 10$, see section \ref{models} for a motivation of this value), but with $\epsilon$ being an essentially free parameter constrained only by 
$\epsilon > \delta/Z_{\rm max}$ (where $Z_{\rm max}$ is the maximum/central metallicity of the galaxy) in order to avoid the singularity that occurs at $Z = \delta/\epsilon$ (see Paper I). In such case the 
dust-to-gas/metals profiles cannot be reproduced for any of the galaxies in our sample. With $\delta$ as a third free parameter (but still $\epsilon > \delta/Z_{\rm max}$),
the best fit is typically obtained for $\delta < 1$.  The only exceptions are NGC 2403 and 3031 for which moderate dust destruction ($\delta\sim 4$) leads to a slight improvement, albeit with
a too high degree of freedom for the model fit to be better from a statistical point of view. This more or less rules out any significant net destruction of dust. Hence, we present here (in detail) only models without
dust destruction due to SN shocks for the selected galaxies, since dust destruction hardly improves the models but rather just adds another free parameter (constraining more the two free 
parameters is also clearly not justified from a statistical point of view, given the small number of data points we have at our disposal for several galaxies). 

\begin{figure*}
  \resizebox{\hsize}{!}{
   \includegraphics{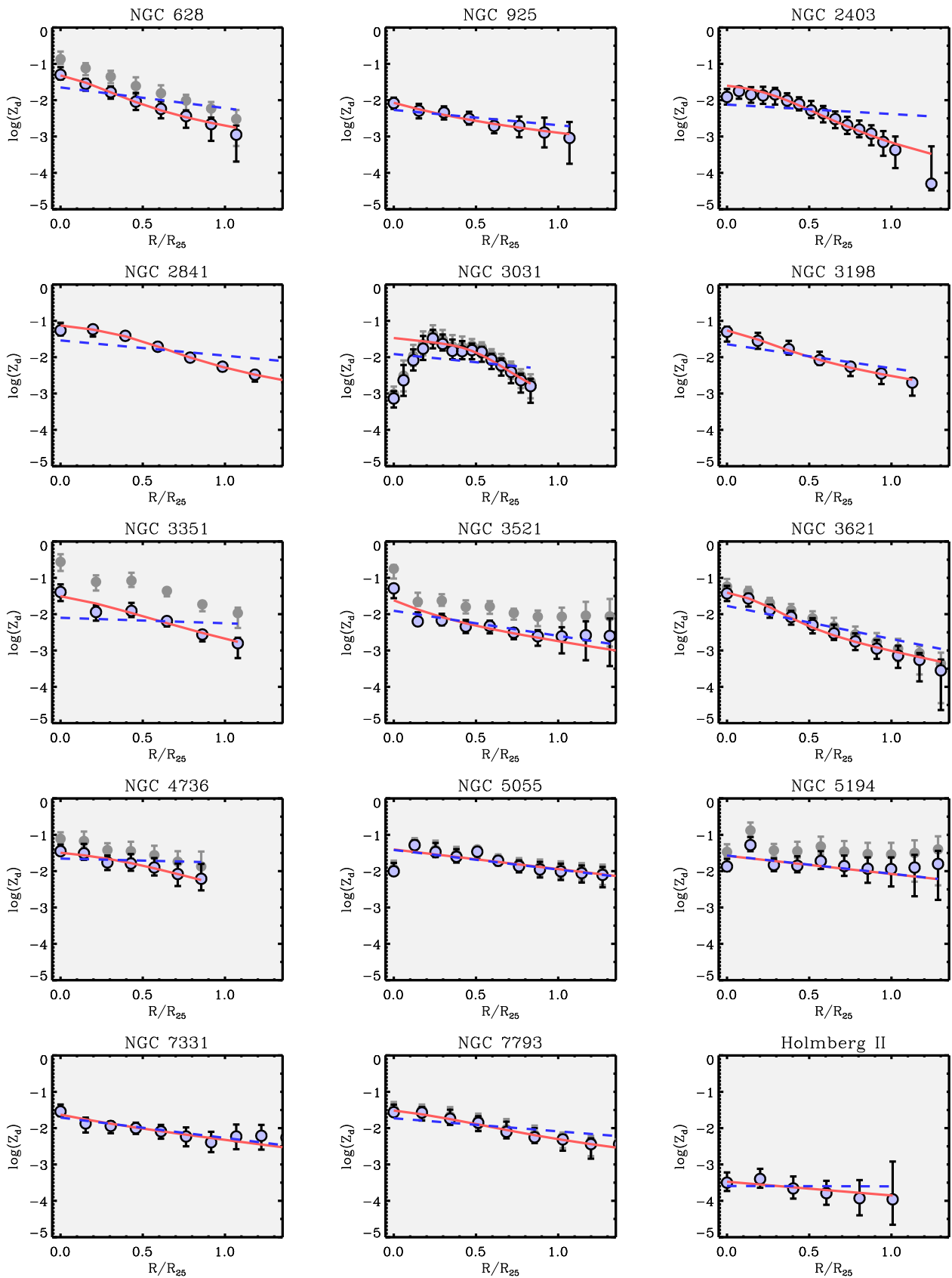}
  }
  \caption{\label{grad} Dust-to-gas ratio as function of galactrocentric distance (circles) together with the best-fit models with stellar dust production only (dashed blue lines) and simple models including dust growth
  (solid red lines). The grey symbols in the background shows the original dust-to-gas ratio before lowering the dust abundance.}
  \end{figure*}

Dust growth in the ISM, on the other hand acts as to improve the overall agreement between models and dust-to-gas/metals profiles derived from observations.  In fact, allowing dust growth  ($\epsilon \ge 0$, 
$\delta = 0$)  improves the fits rather dramatically compared to the case of pure stellar dust production ($\epsilon = \delta = 0$). We find only one case (NGC 3031) where a really good fit cannot be obtained. 
This galaxy show a peculiar central depression in the dust-to-gas profile (see figure \ref{grad}, middle panel on the second row from the top), which may not have
anything to do with the physics of dust formation. A possible scenario is that a merger-induced starburst may have cleansed the central regions from dust, since NGC 3031 appear in a
relatively dense environment (it is for example interacting with NGC 3034 and 3077). For all other galaxies (including the outer part of NGC 3031) we obtain more or less good fits 
(see figure \ref{grad} and figure \ref{zeta}).

In most cases there certainly seem to be need for dust growth, or another kind of 'secondary' dust production, that depends on dust abundance and/or metallicity and thus steepens the dust-to-gas 
gradient and creates a dust-to-metals gradient. The very steep dust-to-metals gradients in the outer parts of NGC 2403 and 3031, as well as that of NGC 3351,
are particularly good examples. In most cases the required stellar dust yield 
$y_{\rm d}$ is well below 50\% of the total metal yield $y_Z$ according to the model fits with $\delta = 0$ and $\epsilon$ being free a parameter (see Table 
\ref{parameters}). This is a very reasonable result, suggesting our model is sound. Models with $\delta = 0$, but with a significant $\epsilon$ and relatively 
modest (or even insignificant) stellar dust production provide very good fits in general  to the dust-to-gas profiles derived by \citet{Munoz-Mateos09}. The cases requiring a high stellar dust yield have high
dust-to-metals ratios over the whole disc (after down-scaling of the dust abundance), suggesting that not only may the dust abundances derived from observations for these galaxies be too high, but also 
the metallicities could be too low. 

In figure \ref{epslogl} we show $\epsilon$ as function of the $B$-band luminosity \citep[data taken from][]{Pilyugin00,Pilyugin04} {and the mean
gas mass density, here estimated as 
\begin{equation}
\langle\Sigma_{\rm g}\rangle \sim {M_{\rm g} \over 2\pi\,R_{25}^{2}},
\end{equation}
\citep[data taken from][]{Pilyugin04,Walter08,Munoz-Mateos09,Moustakas10}}, for models where the fitting does not imply $\epsilon\to 0$ (NGC 
5055, 5194). The $\epsilon$-values correlate nicely with the B-band luminosity suggesting that dust-growth in the ISM is much more efficient in 
smaller and less luminous late-type galaxies. It is unclear why this correlation exists, but we believe there could be a connection to star formation 
and the formation of cold molecular clouds. Low-luminosity disc galaxies  tend to have less star formation which may create a suitable environment 
for dust growth in molecular clouds. If star formation is too intense, dust growth may simply be inhibited by the radiation field. Indirect support for this
hypothesis is that primarily young stars (due to their emission at UV/blue wavelengths) may be responsible for heating cold dust \citep{Gordon04}. 
The $\epsilon$-values does not at all seem to correlate with $\langle\Sigma_{\rm g}\rangle$, however, which suggest the parameterisation we 
introduced in Paper I is very much reasonable.

  \begin{table*}
  \begin{center}
  \caption{\label{parameters} Parameter values for the models. All models have $\delta = 0$.}
  \begin{tabular}{lll|llllllll|l}
  \hline
  Galaxy & Type & ~~~~ & & $\epsilon = 0$ & ~~~~ & & $\epsilon \ge 0$ & ~~~~ & & ~~~~ & Remark\\[2mm]
         & ~~~~ & $y_Z$ & ~~~~ & $y_{\rm d}$ & $y_{\rm d}/y_Z$ & ~~~~ & $y_{\rm d}$ & $y_{\rm d}/y_Z$ & $\epsilon$ & ~~~~ &\\
  \hline
NGC 628  & SAc   & 0.0314 & ~~~~ & 0.0324 & 1.00 & ~~~~ & 0.00322 & 0.103 &       120  & ~~~~ & \\
NGC 925  & SABd  & 0.0156 & ~~~~ & 0.0029 & 0.18 & ~~~~ & 0.00151 & 0.097 &       68.0 & ~~~~ & Interactions with low-mass galaxy?\\
NGC 2403 & SABcd & 0.0188 & ~~~~ & 0.0051 & 0.27 & ~~~~ & 0.00000 & - &       457  & ~~~~ & \\
NGC 2841 & SA(r)b & 0.0292 & ~~~~ & 0.0107 & 0.37 & ~~~~ & 0.00201 & 0.069 &       94.4 & ~~~~ & \\
NGC 3031 & SAab  & 0.0133 & ~~~~ & 0.0047 & 0.35 & ~~~~ & 0.00018 & 0.014 &       216  & ~~~~ & Interacting with NGC 3034, 3077\\
NGC 3198 & SBc   & 0.0148 & ~~~~ & 0.0054 & 0.36 & ~~~~ & 0.00287 & 0.194 &       39.6 & ~~~~ & \\
NGC 3351 & SBb   & 0.0310 & ~~~~ & 0.0315 & 1.00 & ~~~~ & 0.00000 & - &      240  & ~~~~ & Strong bar\\
NGC 3521 & SABbc & 0.0200 & ~~~~ & 0.0150 & 0.75 & ~~~~ & 0.01149 & 0.575 &       24.9 & ~~~~ & \\
NGC 3621 & SAd?  & 0.0166 & ~~~~ & 0.0105 & 0.63 & ~~~~ & 0.00376 & 0.227 &       177  & ~~~~ & Pure disc/bulgeless galaxy\\
NGC 4736 & SA(r)ab & 0.0119 & ~~~~ & 0.0069 & 0.56 & ~~~~ & 0.00001 &  0.001 &      152  & ~~~~ & \\
NGC 5055 & SAbc  & 0.0306 & ~~~~ & 0.0189 & 0.62 & ~~~~ & 0.01825 & 0.596 &       0.00 & ~~~~ & Interacting with UGC 8365?\\
NGC 5194 & SABbc & 0.0247 & ~~~~ & 0.0217 & 0.88 & ~~~~ & 0.02470 & 1.000 &       0.00 & ~~~~ & Interacting with NGC 5195\\
NGC 7331 & SAb   & 0.0190 & ~~~~ & 0.0068 & 0.36 & ~~~~ & 0.00586 & 0.308 &       9.94 & ~~~~ & Retrograde bulge\\
NGC 7793 & SAd   & 0.0144 & ~~~~ & 0.0092 & 0.64 & ~~~~ & 0.00284 & 0.197 &       140  & ~~~~ & \\
Holmberg II & dIrr & 0.0086 & ~~~~ & 0.0005 & 0.05 & ~~~~ & {\it 0.00000} & - &    {\it 14018}  & ~~~~ & No measured metallicity slope\\  
  \hline
  \end{tabular}
  \end{center}
  \end{table*}

\subsection{Comments on models of individual galaxies}
Here we summarise and comment on the modelling results of individual galaxies.

\subsubsection{NGC 628} 
The dust-to-metals gradient is clearly negative. Assuming pure stellar dust production implies that the stellar dust yield $y_{\rm d}$ must be equal to the total metal yield $y_Z$ 
in order to match the dust abundance in general, which is obviously unrealistic (especially since the dust abundance has been scaled down by 62\%). Adding 
significant dust growth in the ISM provides a good model with $y_{\rm d}$ being only about 10\% of $y_Z$. A fairly strong indication of non-stellar dust production.

\subsubsection{NGC 925}  
This galaxy also show a negative dust-to-metals gradient, but the slope is shallower. The dust-to-gas/metals gradient can be nicely modelled with dust growth.
May be (or have been) interacting with a low-mass galaxy, which could explain dynamical and morphological asymmetries \citep{Pisano98}.

\subsubsection{NGC 2403} 
This is a somewhat peculiar case. The inner part of the dust-to-gas gradient has a slope which is very much consistent with the slope of the metallicity
gradient, while the outer part is significantly steeper. Beyond $R/R_{25}\sim 0.3$ the observationally inferred dust-to-metals profile show a steep drop which is
clearly inconsistent with the model with only stellar dust production. In that case, the model prediction falls outside the error bars for $R/R_{25}\geq 0.7$. 

Adding dust growth improves the agreement between model and observation tremendously. This is one of the strongest cases in support of grain growth
being the dominant mechanism for dust production in the ISM. However, the very inefficient stellar dust production ($y_{\rm d}$ only $\sim 1$\% of $y_Z$) 
suggested by the model fit to the data is indicative of a possibly {\it too low} dust-to-metals ratio in the outer disc. The ratio is less than 1\% in the outskirts of the disc and 
$\sim 50$\% in the inner disc (see figure \ref{zeta}).

\subsubsection{NGC 2841}
The dust-to-metals gradient is quite steep, but flattens out towards the centre of the galaxy. Both features can be perfectly reproduced by a model including dust growth 
and the required/implied stellar dust yield is just 7\% of the metal yield, which makes this galaxy another strong case for dust growth - in particular since the error bars
are quite small in this case. A model fit with only stellar dust production is clearly inconsistent with the dust-to-gas/metals gradient. The dust abundance was only scaled 
down by 5\% in this case.

\subsubsection{NGC 3031} 
NGC 3031(M 81) is another peculiar case, where there is a clear dust depletion in the central parts. This is certainly not predicted by the model. The mid part of
the disc has a dust-to-metals gradient which is nearly flat, but beyond $R/R_{25}\sim 0.6$ it steepens in a way similar to NGC 2403.
It is worth noting that this galaxy is known to be in a relatively dense environment and interact with at least two other galaxies \citep{Davidge08}, which suggest 
that the central parts likely have formed by one or more mergers. A merger is a violent event, quite capable of destroying large amounts of dust, which is a
possible explanation of the central depression in the dust-to-gas profile. SN-shock waves may not be efficient dust destroyers, although the environment
in merger-driven nuclear starburst may be hostile enough to actually destroy dust grains and not just crush them into smaller grains.
A model with significant dust growth in the ISM, but almost no stellar dust production, reproduces the slope of the outer dust-to-gas/metals profile nicely. Again, we
argue this is strong evidence for grain growth in the ISM being the most important dust formation mechanism. The dust abundance was scaled down by 25\%.

\subsubsection{NGC 3198} 
Once again the dust-to-metals gradient is clearly negative. Assuming pure stellar dust
production is quite reasonable in this case ($y_{\rm d}$ being about 36\% of $y_Z$), although including dust growth enhances the agreement between model and
observation significantly. Stellar dust production is still relatively important in this case, where $y_{\rm d}$ is about 19\% of $y_Z$.

\subsubsection{NGC 3351} 
This case is interesting, since it shows the largest discrepancy between the distributions of metals and dust. The metallicity gradient is almost flat allover
the disc. A model without dust growth is ruled  out for two reasons: pure stellar dust production implies $y_{\rm d} = y_Z$, and the corresponding dust-to-gas/metals 
profile does not fit the data at all. Adding dust growth, however, we obtain a clearly better fit to the observations (see figure \ref{grad} and figure \ref{zeta}). 
Yet another case in favour of somewhat lower dust abundances. To ensure the dust-to-metals ratio never exceeds 0.9, the dust abundance had to be scaled down 
by a massive 85\%, which indicate there might be a serious problem with the method \citep[see][]{Draine07,Munoz-Mateos09} for deriving dust abundances from 
observations.

\subsubsection{NGC 3521} 
The model with pure stellar dust production fits quite nicely to the observational dust-do-gas/metals profile, albeit with a rather large stellar dust yield 
($y_{\rm d}$ being about 75\% of $y_Z$) and with an exception for the inner most data point. With dust growth, the inner most point is still not reproduced,
so the agreement with the observations is not clearly better (see figure \ref{grad} and figure \ref{zeta}). The dust abundance was scaled down by 71\%, which again
provide evidence that there is a problem with how dust abundances are derived.

\subsubsection{NGC 3621} 
The dust-to-metals gradient have clearly a negative slope, but the disagreement with a pure stellar dust production is moderate. Dust growth improves the overall 
agreement between model and observation, and an essentially perfect fit is obtained for very reasonable parameter values. The model clearly reproduces the 
dust-to-gas as well as the dust-to-metals profile within the error bars. The dust abundance was scaled down by 35\%.

NGC 3621 is considered to be a pure disc (bulgeless) galaxy, which imply a quiescent formation history without major mergers and that central regions are
fairly unevolved compared to other late-type galaxies \citep{Gliozzi09}. Hence, this galaxy may be a good test case for an idealised model such as the one we
have applied here.

\subsubsection{NGC 4736}
This galaxy has a clear dust-to-metals gradient, which is nicely reproduced by a model with significant dust growth. Stellar dust production is essentially rejected
by the fitting algorithm ($y_{\rm d} \sim 10^{-5}$), which is unphysical, although it implies dust growth is needed and clearly dominating. A model fit with only stellar 
dust production is inconsistent with the dust-to-gas gradient, but almost within the error bars, which makes it marginally consistent with the dust-to-metals gradient
since the error bars are larger for dust-to-metals ratios. However, a model with dust growth is clearly preferred. The dust abundance was scaled down by 54\%,
which yet again seems problematic.

\subsubsection{NGC 5055} 
\begin{figure*}
  \resizebox{\hsize}{!}{
   \includegraphics{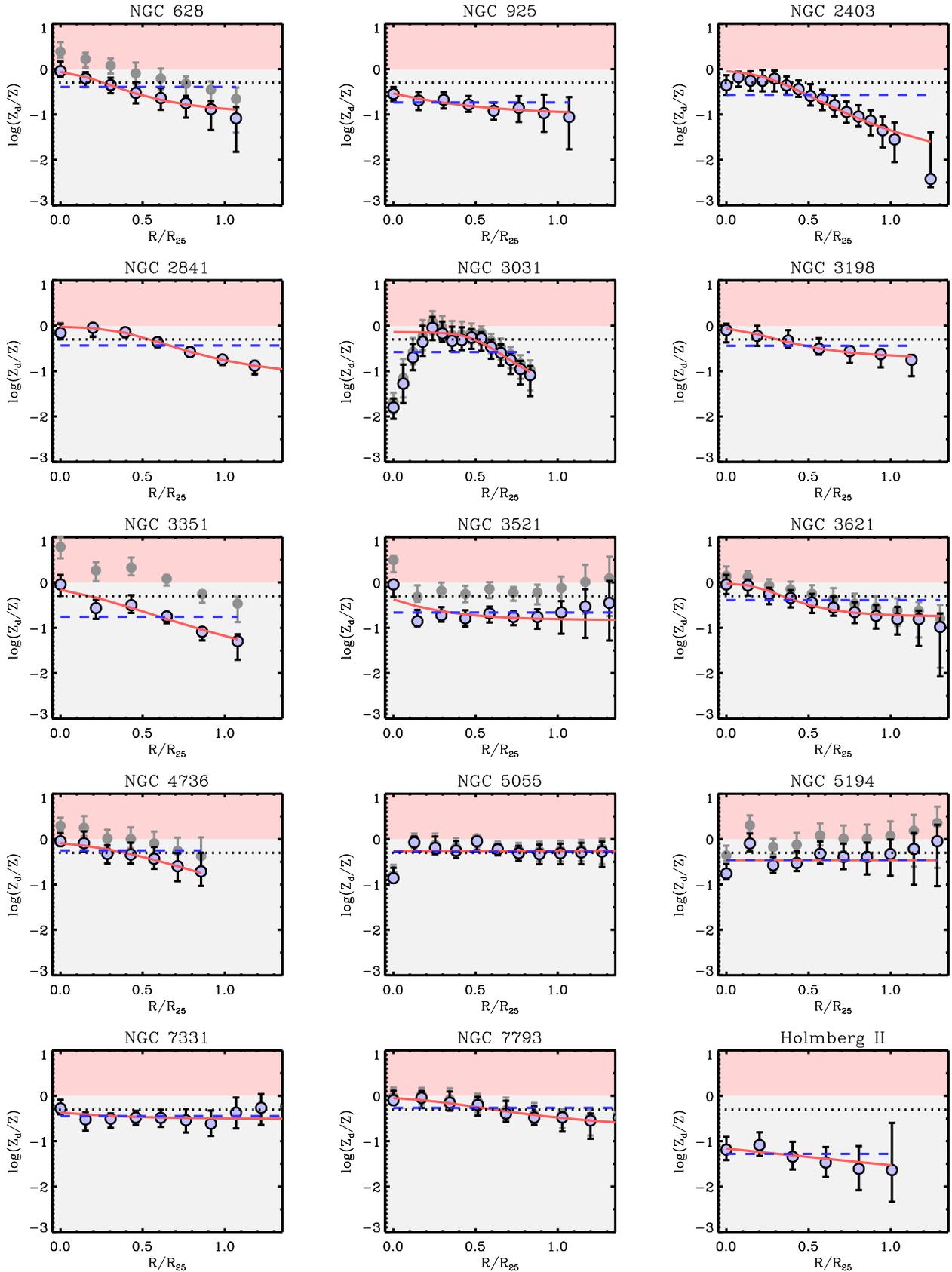}
  }
  \caption{\label{zeta} Dust-to-metal ratio as function of galactrocentric distance (circles) together with the best-fit models with stellar dust production only (dashed blue lines) and simple models including dust growth 
  (solid red lines). The dotted (black) line shows the case of 50\% metals locked-up in dust, which roughly corresponds to the dust-to-metals ratio in the Solar neighbourhood.
  The grey symbols in the background shows the original dust-to-metals ratio before lowering the dust abundance. {The light-red/pink shaded regions correspond to dust-to-metals ratios above unity.}}
  \end{figure*}

Here, the dust-to-metals gradient is almost exactly flat (see figure \ref{zeta}). As a consequence the fitting routine reject the dust growth contribution
($\epsilon\to 0$). The required stellar dust yield $y_{\rm d}$ is 62\% of the metal yield $y_Z$, which is not a totally unreasonable figure. The inner most data point
suggest a mild central depletion in the dust-to-gas profile. NGC 5055 is possibly interacting with UGC 8365. The dust abundance was scaled down by 14\%.

\subsubsection{NGC 5194} 
NGC 5194 is the only galaxy in the present study which deviates from the general trend that dust-to-metals gradients have negative slopes.  In this galaxy the 
dust-to-gas profile is in principle flat over the whole disc, while the dust-to-metals gradient shows a mildly positive slope. Hence, NGC 5194 could seem better 
modelled with some moderate dust destruction, but dust destruction would require a higher stellar dust yield, and since the model with only stellar dust production 
suggest that 88\% of the metals ejected from stars is in the form of dust, there is not much room for dust destruction. Dust growth means no improvement of the fit in this case.
This galaxy show unrealistic dust-to-metals ratios at several locations along the disc, without down-scaling of the dust abundance. The dust abundance was scaled down 
by 60\%, which was not enough to bring down the stellar dust yield to a reasonable number, however. This galaxy is somewhat unusual in that it appears to have an
essentially flat metallicity gradient, something which is not expected for this morphological type (SABbc). The need for a 60\% down-scaling of the dust abundance 
would not be if the metallicity of the central parts were as much higher as needed to create a more typical metallicity gradient.

     \begin{figure*}
  \resizebox{\hsize}{!}{
   \includegraphics{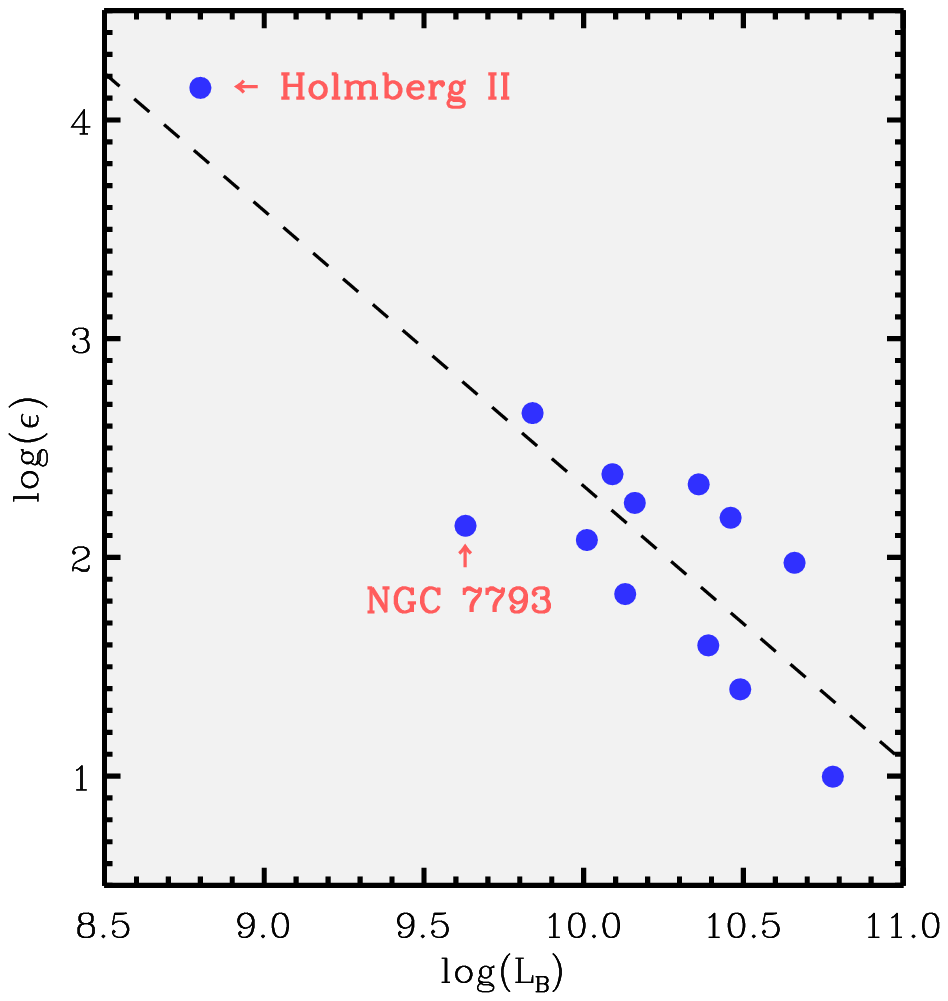}
   \includegraphics{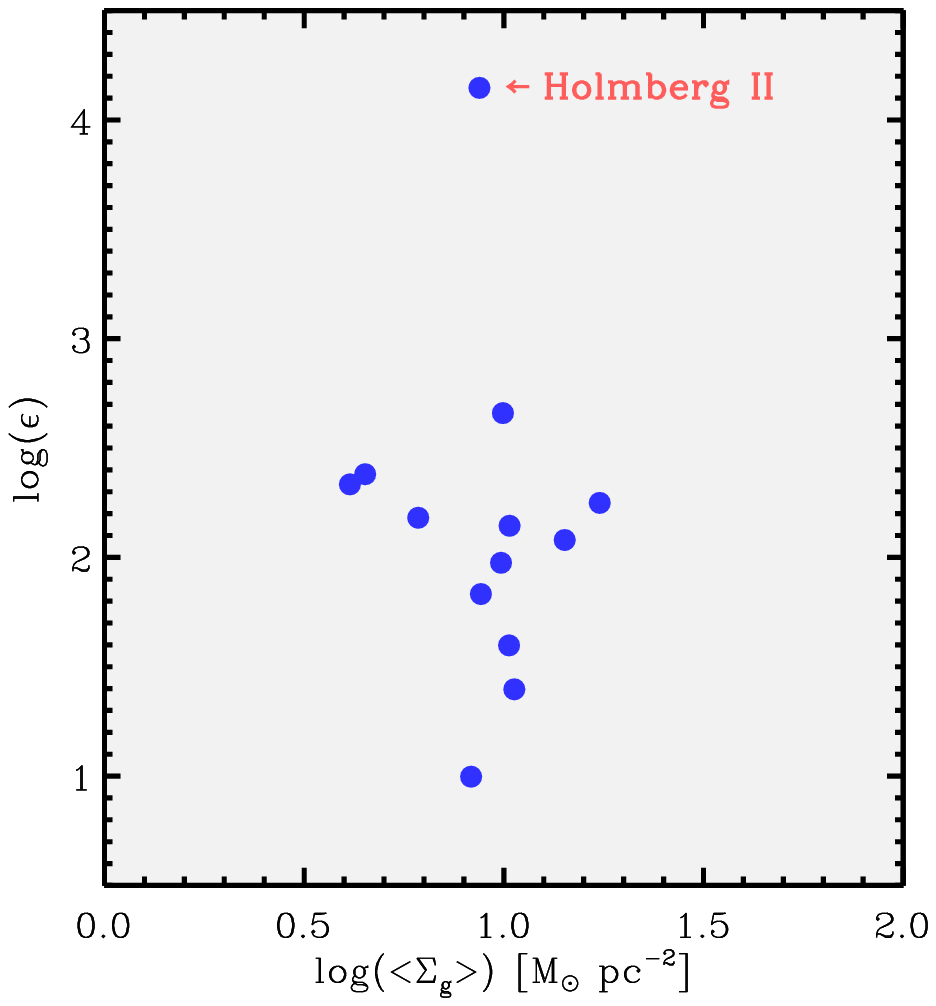}
  }
  \caption{\label{epslogl} The correlation between the dust-growth parameter $\epsilon$ (for models where the fitting does not imply $\epsilon\to 0$) and the B-band luminosity  {(left) and average gas mass density (right)} of each galaxy.}
  \end{figure*} 

\subsubsection{NGC 7331} 
The dust-to-metals gradient is again nearly flat. Adding dust growth means only a minor improvement and the stellar dust yield required in the
pure stellar dust model is very much reasonable ($y_{\rm d}$ being about 36\% of $y_Z$). NGC 7331 is therefore an example of that there is not always an obvious
need for dust growth. A remarkable property of this galaxy is its retrograde bulge (counter-rotating relative to the disc), a fact which is not likely to be connected to the
distributions of metals and dust.

\subsubsection{NGC 7793} 
This galaxy shows a moderately negative dust-to-metals gradient. A model with only stellar dust production works reasonably
well, although it is not totally within the error bars. Adding dust growth yields an excellent model fit. The dust abundance was scaled down by 15\%.

\subsubsection{Holmberg II}
This galaxy is the only one in our sample which is not really a disc galaxy. It is a dwarf irregular galaxy, which seems to have an essentially flat 
metallicity gradient and generally a low dust abundance, as is usually the case with galaxies of this morphological type. There is a dust-to-metals 
gradient, however, which can be nicely reproduced by a model with dust growth, provided the metallicity gradient is not exactly flat. Here we have 
(arbitrarily)  assumed a 1\% slope, which gives a reasonable result {(although the high $\epsilon$-value makes Holmberg II deviate from the 
overall location of the data points in the right panel of figure \ref{epslogl})}. The required/implied stellar dust yield is approaching zero 
($y_{\rm d}\to 0$), suggesting dust  growth must be the dominating process for dust production. However, it should be noted that a model with only 
stellar dust production is in principle  consistent (within the errors) of the dust-to-gas/metals gradients derived from observations. The stellar dust 
yield would in such case need to be only 5\% of the metal yield.

\subsection{If dust is not destroyed - then what is the effect of shocks from SNe?}
Clearly, our results presented above suggest dust is not destroyed in any significant quantities due to SN shocks, as have been suggested by several studies and authors
\citep[see, e.g.][]{Draine79,McKee89,Tielens94,Nozawa06,Jones96,Jones04,Jones11}. We propose dust {\it reprocessing} rather than destruction, which we motivate as follows. 

The first thing to consider is what we mean by 'dust destruction'. Here, we prefer to look at dust destruction as a process that will {atomise} a dust grain/molecule. A SN-shock wave contains 
a vast amount of energy, which can certainly be harmful for dust grains, but it is quite possible that dust grains
are shattered to small splinters, rather than completely evapourated. Small grains tend to survive better when exposed to energetic shock waves \citep[see, e.g.][]{Jones96},
which implies there might be a lower size limit at which the shattering stops being effective. Such a scenario has the advantage of being favourble to dust growth in the ISM. Assuming spherical
grains {the total surface area $A_{\rm tot}$ of all dust grains is related to the total volume $V_{\rm tot}$ of dust as $A_{\rm tot}= 3V_{\rm tot}/\langle a\rangle$, where $\langle a\rangle$ is the  
the average grain radius. Thus, if the average grain size is decreases by a certain factor, the total surface area $A_{\rm tot}$ increases by the same factor.}
The dust-growth timescale then decreases with a similar factor (see equation \ref{taugr}). Recent, more detailed, analysis of this fact is presented by \citet{Jones11} and \citet{Hirashita11}.
In the latter study, \citet{Hirashita11} show that the time-scale of the grain growth is very sensitive to the grain-size distribution, because the grain growth is mainly regulated by the 
surface-to-volume ratio of grains, as mentioned above.

This 'grain-shattering 
scenario' also helps to avoid a too large population of extremely large grains in the ISM, as recent developments in modelling of dust growth and wind formation in AGB-star atmospheres 
suggest dust grains may be relatively large (micron-sized) already when they leave these stars \citep{Hoefner08,Mattsson11b}. Attempts to solve the inverse-problem of finding the grain-size
distribution from interstellar extinction measurements suggest dust grains are typically not larger than $\sim 1\mu$m \citep[see, e.g.,][]{Clayton03}, which would pretty much rule out
the existence of extremely large grains in the ISM. AGB stars contribute a significant fraction of stellar dust, so the apparent absence of grains larger than $\sim 1\mu$m must either suggest
that grains are both grown and shattered to smaller fragments in the ISM, or the micron-sized grains predicted by dynamic AGB-star atmosphere modelling is a false prediction.

Dust grains may also have their chemical-bond structure altered due to external forces rather than being shattered or destroyed \citep[see, e.g.][]{Tielens87,Jones90,Vollmer07}. For example, 
production of graphite-like carbon dust and even nano-diamonds, is a likely outcome of shock-processing of interstellar dust. This type of dust re-processing is desirable since it is unlikely
stars can produce the amount of graphite (or graphite-like) material needed to explain the 2175\AA -bump in, e.g., the Milky Way extinction curve \citep{Mathis77,Pei92, Kim94, Clayton03}.
Carbonaceous dust may also produce hydrocarbons when exposed to shock waves \citep{Taylor93}, which would be desirable for similar reasons, as hydrocarbons may contribute to the
2175\AA -feature \citep{Blanco91} as well as explain infrared features \citep{Kwok11}. 

From a more general point of view, there is evidence for dust reprocessing in SN remnants from differences in dust species ratios. In a recent study, \citet{Andersen11} found the ratio of very 
small grains to big grains to be higher than that found in the plane of the Milky Way in a study of molecular interacting SN remnants. The discrepancy is typically a factor of 2Ð3. They suggest 
that dust shattering is responsible for the relative over-abundance of small grains, which is consistent with the grain-shattering scenario we suggest and in agreement with prediction from dust
destruction models \citep{Jones11}. In some cases small carbon grains are severely underabundant, which is likely evidence for sputtering, according to \citet{Andersen11}.

\subsection{Are dust abundances generally overestimated?}
\label{dustestimates}
As it was pointed out above, we believe the dust abundance may be overestimated in a several (if not all) cases. \citet{Munoz-Mateos09} uses the model by \citet{Draine07} which
assumes a specific composition of different dust species based on the inferred dust properties of the solar neighbourhood (except the variations of the PAH abundance) 
and a fixed grain-size distribution. We believe this to be a too simple model of the dust component in a galaxy, mainly because the dust properties in the local Galaxy most 
certainly do not reflect the dust properties everywhere else and that the grain-size distribution most likely changes with environment \citep{Hirashita11}. 

We would also like to point out that the dust-to-hydrogen ratio suggested for the local Galaxy by \citet{Draine07b} , i.e., $M_{\rm d}/M_{\rm H} = 0.010$, is similar
to the metals-to-hydrogen ratio for the solar neighbourhood (essentially the same as in the Sun), which implies a dust-to-metals ratio {of $\sim 0.6$. Observationally, this ratio is}
$\sim 0.3 - 0.5$ \citep{Whittet91} which is also roughly the ratio found in most local galaxies of similar type \citep[see, e.g.][]{Issa90}. {Hence, there is a possibility that} the dust abundance 
could be overestimated {due to the model itself}. Evidence in support of this idea has recently been presented by \citet{Madden11}, suggesting that the dust-to-gas ratios of dwarf galaxies 
are not in accord with the low metallicity of these systems.

Furthermore, the estimates by \citet{Draine07} uses a total carbon abundance of $245 \pm 30$~ppm relative to H from \citet{Asplund05}. However, \citet{Landi07} and \citet{Centeno08} obtain values of 
(O/H)$_{\sun}$ which are $\sim 1.9$ times larger than the solar oxygen value of (O/H)$_{\sun} = 457 \pm 56$ppm from \citet{Asplund05}. The solar carbon abundance might therefore be significantly larger than 
the 245ppm used by \citet{Draine07}. If the total carbon abundance is rather 350~ppm, the mass of carbon in dust would {\it increase} by 75\% according to \citet{Shen08}. 
{Note, however, that \citet{Compiegne11} also obtain $M_{\rm d}/M_{\rm H} = 0.010$ using a different model than that of \citet{Draine07}, which suggest a total carbon abundance of $269 \pm 33$~ppm relative to H 
\citep[from][]{Asplund09} where 26\% is gas phase carbon \citep[to be compared to the 4\% in the model by][]{Draine07}. }

Even if the dust abundances may need to be scaled down, we believe the errors are systematic in a way which will not affect the actual slope of the dust-to-gas
profile very much. Hence, we consider the results we have obtained here being most likely qualitatively correct, i.e., the dust-to-gas gradients are indeed steeper
than the metallicity gradients, suggesting significant dust growth in the ISM.

\section{Conclusions}
Dust destruction by shocks from exploding SNe is believed to be a rather efficient process, evaporating dust grains such that large volumes of surrounding interstellar space is essentially cleared from dust. We 
challenge this viewpoint, and suggest that SN shocks induce {\it reprocessing} of interstellar dust rather than 
destroying it. We base our conclusion on the fact that the effect of dust destruction by SN-shock waves on the dust-to-metals gradients in late-type galaxies is not compatible with these gradients being negative
in general, which is what observations tell us at this point. SN-shocks may indeed still crush dust grains into small splinters, thus changing the grain-size distribution, but perhaps not disintegrate
dust grains completely. Moreover, dust grains may change properties in other ways as well, e.g., through annealing or restructuring of the chemical bonds. The latter may be a way to produce graphite-like dust
in the ISM - a type of carbon dust which is unlikely to be formed in any larger quantities in stellar atmospheres.

Two other results also stand out: (1) dust growth in the ISM seem to be the most important dust formation mechanism, and (2) conventional methods for estimating dust abundances may include systematic 
errors which are not fully under control. Assuming stellar dust production only, we found that dust-to-gas gradients are generally too steep to be reproduced by such a model. Dust growth is an efficient way to
steepen the dust-to-gas gradient relative to the metallicity gradient, which makes it an inevitable ingredient in any attempt to model the distribution of dust along galactic discs. In many cases we found the dust
growth to be the totally dominant dust production mechanism. However, the need for dust growth may be slightly exaggerated in our models, since there are cases suggesting the dust abundances
derived from observations are too high, even after the dust abundances have been scaled down. In these cases the dust-to-metals ratio (as obtained from observations) exceeds unity without rescaling, which is
clearly unphysical. If the amount of dust is systematically overestimated, this may of course affect the need for dust growth. 

Regardless of whether the dust abundances should be lower or not, the metal content of the inner part of the galaxies seem dominated by dust in most cases. Even if the dust-to-metals ratios suggest the dust 
abundances may be overestimated in the central regions in particular, there is a general trend towards higher dust-to-metals at low galactocentric radii (a natural consequence of the dust-to-gas gradient being
steeper than the metallicity gradient). This suggest the extinction properties may vary significantly over  the disc of a late-type galaxy and the amount of extinction may not simply scale with metallicity. If so, 
corrections for extinction should not be made using a generic correction method applied to the whole disc.
 
\section*{Acknowledgments}
The authors thank the reviewer, Anthony Jones, for constructive and helpful comments and criticism that greatly helped to improve the presentation.
L.M. acknowledges support from the Swedish Research Council (Vetenskapsr\aa det). The Dark Cosmology Centre is funded by the Danish National Research Foundation.

This research has made use of the NASA/IPAC Extragalactic Database (NED) which is operated by the Jet Propulsion Laboratory, California Institute of Technology, under contract with the National Aeronautics
and Space Administration.


\begin{thebibliography}{}
\bibitem[\protect\citeauthoryear{Andersen et al.}{2011}]{Andersen11}
  Andersen M., Rho J., Reach W. T., Hewitt J. W. \& Bernard J. P., 2011, ApJ, accpeted
\bibitem[\protect\citeauthoryear{Asplund et al.}{2009}]{Asplund09}
  Asplund M., Grevesse N., Sauval A.J. \& Scott P., 2009, ARA\&A, 47, 481
\bibitem[\protect\citeauthoryear{Asplund et al.}{2005}]{Asplund05}
  Asplund M., Grevesse N. \& Sauval, A.J., 2005, ASPC, 336, 25A
\bibitem[\protect\citeauthoryear{Bianchi \& Schneider}{2007}]{Bianchi07}  
  Bianchi S. \& Schneider R., 2007, MNRAS 378, 973
\bibitem[\protect\citeauthoryear{Bigiel et al.}{2008}]{Bigiel08}    
  Bigiel, F., Leroy, A., Walter, F., Brinks, E., de Blok, W. J. G., Madore, B., \& Thornley, M. D. 2008, AJ, 136, 2846
\bibitem[\protect\citeauthoryear{Bigiel et al.}{2011}]{Bigiel11}      
  Bigiel F., Leroy A. K., Walter F., Brinks E., de Blok W. J. G., Kramer C., Rix H. W., Schruba A. et al, 2011, ApJ, 730, L13+
\bibitem[\protect\citeauthoryear{Blanco et al.}{1991}]{Blanco91}        
  Blanco A., Bussoletti E., Colangelli L., Fonti S. \& Stephens J. R., 1991, ApJ, 382, 97
\bibitem[\protect\citeauthoryear{Centeno \& Socas-Navarro}{2008}]{Centeno08}   
  Centeno R. \& Socas-Navarro H., 2008, ApJ, 682m L61
\bibitem[\protect\citeauthoryear{Clayton}{1987}]{Clayton87}   
  Clayton, D. D. 1987, ApJ, 315, 451
\bibitem[\protect\citeauthoryear{Clayton et al. }{2003}]{Clayton03}   
  Clayton, G. C., Wolff M. J., Sofia U. J., Gordon K. D. \& Misselt K. A., 2003, ApJ, 588, 871
\bibitem[\protect\citeauthoryear{Compi\`egne et al. }{2011}]{Compiegne11}   
  Compi\`egne M., Verstraete L., Jones A., et al., 2011, A\&A, 525, A103
\bibitem[\protect\citeauthoryear{Davidge}{2008}]{Davidge08}  
   Davidge T. J., 2008, PASP, 120, 1145
\bibitem[\protect\citeauthoryear{Draine \& Salpeter}{1979}]{Draine79}  
  Draine B. T. \& Salpeter E. E., 1979, ApJ, 231, 438
\bibitem[\protect\citeauthoryear{Draine}{1990}]{Draine90}
  Draine B., 1990, in BlitzL., ed., ASP Conf. Ser. Vol. 12, 
  The Evolution of the Interstellar Medium. Astron. Soc. Pac., San Francisco , p. 193
\bibitem[\protect\citeauthoryear{Draine et al.}{2007}]{Draine07b}  
  Draine B. T., Dale D. A., Bendo G., et al., 2007, ApJ, 663, 866 
\bibitem[\protect\citeauthoryear{Draine \& Li}{2007}]{Draine07}  
  Draine B. T. \& Li A. 2007, ApJ, 657, 810
\bibitem[\protect\citeauthoryear{Dunne et al.}{2009}]{Dunne09}
  Dunne L., Maddox S. J., Ivison R. J., et al. 2009, MNRAS, 394, 1307
  \bibitem[\protect\citeauthoryear{Dwek}{1998}]{Dwek98} 
  Dwek E., 1998, ApJ, 501, 643
\bibitem[\protect\citeauthoryear{Dwek et al.}{2007}]{Dwek07} 
  Dwek E., Galliano F. \& Jones A.P., 2007, ApJ, 662, 927
\bibitem[\protect\citeauthoryear{Edmunds}{1990}]{Edmunds90}
  Edmunds M.G., 1990, MNRAS, 246, 678   
\bibitem[\protect\citeauthoryear{Edmunds \&  Eales}{1998}]{Edmunds98}
  Edmunds M.G. \& Eales S.A., 1998, MNRAS, 299, L29
\bibitem[\protect\citeauthoryear{Edmunds}{2001}]{Edmunds01}
  Edmunds M.G., 2001, MNRAS, 328, 223  
\bibitem[\protect\citeauthoryear{Feldmann, Gnedin \& Kravtsov }{2011}]{Feldmann11} 
  Feldmann R., Gnedin N. Y. \& Kravtsov A. V., 2011, ApJ, 732, 115
\bibitem[\protect\citeauthoryear{Ferrarotti \& Gail}{2006}]{Ferrarotti06}  
  Ferrarotti A.S. \& Gail H.-P., 2006, A\&A, 447, 553
\bibitem[\protect\citeauthoryear{Gall, Andersen \& Hjorth}{2011}]{Gall11}
  Gall C., Andersen A. C. \& Hjorth J., 2011a, A\&A, 528, A13
\bibitem[\protect\citeauthoryear{Garnett et al.}{2002}]{Garnett02}
  Garnett, D.R., 2002, ApJ, 581, 1019
\bibitem[\protect\citeauthoryear{Garnett}{1997}]{Garnett97}
  Garnett, D.R., Shields G. A., Skillman E. D., Sagan S. P. \& Dufour R. J., 1997, ApJ, 489, 63
\bibitem[\protect\citeauthoryear{Gliozzi et al.}{2009}]{Gliozzi09}  
  Gliozzi M., Satyapal S., Eracleous M., Titarchuk L., \& Cheung C. C., 2009, ApJ, 700, 1759
\bibitem[\protect\citeauthoryear{Gordon et al.}{2004}]{Gordon04}    
  Gordon K. D., P\'erez-Gonz\'alez, Misselt K. A., et al., 2004, ApJSS, 154, 215
\bibitem[\protect\citeauthoryear{Hirashita}{1999}]{Hirashita99}  
  Hirashita H., 1999, ApJ, 510, L99
\bibitem[\protect\citeauthoryear{Hirashita \& Kuo}{2011}]{Hirashita11}    
  Hirashita H. \& Kuo T.-M., 2011, MNRAS, accepted
\bibitem[\protect\citeauthoryear{H\"ofner}{2008}]{Hoefner08}   
  H\"ofner S., 2008, A\&A, 491, L1
\bibitem[\protect\citeauthoryear{Inoue}{2003}]{Inoue03}  
  Inoue A. K., 2003, PASJ, 55, 901
\bibitem[\protect\citeauthoryear{Issa et al.}{1990}]{Issa90}
  Issa M.R., MacLaren I. \& Wolfendale A.W., 1990, A\&A, 236, 237
\bibitem[\protect\citeauthoryear{Jones}{2004}]{Jones04}  
  Jones, A. P. 2004, in ASP Conf. Ser. 309, Astrophysics of Dust, ed. A. N. Witt,
  G. C. Clayton, \& B. T. Draine (San Francisco: ASP), 347
\bibitem[\protect\citeauthoryear{Jones, Duley \& Williams}{1990}]{Jones90}   
  Jones A. P., Duley W. W. \& Williams D. A., 1990, QJRAS, 31, 567
\bibitem[\protect\citeauthoryear{Jones \& Nuth}{2011}]{Jones11}   
  Jones A. P. \& Nuth J. A., 2011, A\&A, 530, A44
\bibitem[\protect\citeauthoryear{Jones, Tielens \& Hollenbach}{1996}]{Jones96}   
  Jones A. P., Tielens A. G. G. M. \& Hollenbach D. J., 1996, ApJ, 469, 740
\bibitem[\protect\citeauthoryear{Kennicutt et al.}{2003}]{Kennicutt03}     
  Kennicutt, R. C. Jr., et al., 2003, PASP, 115, 928
\bibitem[\protect\citeauthoryear{Kim, Martin \& Hendry}{1994}]{Kim94} 
 Kim S.-H., Martin P. G., \& Hendry P. D., 1994, 422, 164
\bibitem[\protect\citeauthoryear{Kobulnicky \& Kewley}{2004}]{Kobulnicky04}      
  Kobulnicky H. A. \& Kewley L. J., 2004, ApJ, 617, 240
\bibitem[\protect\citeauthoryear{Kotak et al.}{2006}]{Kotak06}
  Kotak R., Meikle P., Pozzo M., et al., 2006, ApJ, 651, L117
\bibitem[\protect\citeauthoryear{Kotak et al.}{2009}]{Kotak09}    
  Kotak R., Meikle W.P.S., Farrah D., et al., 2009, ApJ, 704, 306
\bibitem[\protect\citeauthoryear{Krumholz, McKee \& Tumlinson}{2009}]{Krumholz09}     
  Krumholz M. R., McKee C. F. \& Tumlinson J., 2009, ApJ, 699, 850
\bibitem[\protect\citeauthoryear{Krumholz, Leroy \& McKee}{2011}]{Krumholz11}   
  Krumholz M. R., Leroy A. K. \& McKee C. F., 2011, ApJ, 731, 25
\bibitem[\protect\citeauthoryear{Kummer}{1837}]{Kummer1837}   
  Kummer E. E., 1837, Journal f\"ur die reine und angewandte Mathematik, 17, 228
\bibitem[\protect\citeauthoryear{Kwok \& Zhang}{2011}]{Kwok11}   
  Kwok S. \& Zhang Y., 2011, Nature, ??, ??
\bibitem[\protect\citeauthoryear{Landi, Feldman \& Doschek}{2007}]{Landi07}
  Landi E., Feldman U. \& Doschek G. A., 2007, ApJ, 659, 743
\bibitem[\protect\citeauthoryear{Larson}{1998}]{Larson98}
  Larson R.~B. 1998, MNRAS, 301, 569
\bibitem[\protect\citeauthoryear{Leroy et al.}{2008}]{Leroy08}
  Leroy A. K., Walter F.,  Brinks E.,  Bigiel F., de Blok W. J. G.,  Madore B., Thornley M.D. \& Adam Leroy, 2008, AJ, 136, 2782
\bibitem[\protect\citeauthoryear{Madden et al.}{2011}]{Madden11}    
  Madden S. C. Maud Galametz M., Cormier D., et al., 2011, to appear in the proceedings of the 5th Zermatt Symposium: "Conditions 
  and Impact of Star Formation: New Results from Herschel and Beyond" September 19-24, 2010
\bibitem[\protect\citeauthoryear{Magrini et al.}{2011}]{Magrini11}  
  Magrini L., Bianchi S., Corbelli E., et al., 2011, A\&A, accepted, arXiv:1106.0618
\bibitem[Markwardt (2009)]{Markwardt09}
  Markwardt C.B., 2009, in Astronomical Society of the Pacific Conference Series, Vol. 411, D. Bohlender, D. Durand, \& P. Dowler, ed.,  
  Data Analysis Software and Systems XVIII, ASP Conference Series, p.251
\bibitem[\protect\citeauthoryear{Mathis, Rumpl \& Nordsieck}{1977}]{Mathis77}  
  Mathis J. S., Rumpl W. \& Nordsieck K. H., 1977, ApJ, 217, 425
\bibitem[\protect\citeauthoryear{Matsuura et al.}{2011}]{Matsuura11}
  Matsuura M., Dwek E., Meixner M., et al., 2011, Science, 333, 1258
\bibitem[\protect\citeauthoryear{Mattsson}{2011}]{Mattsson11}
  Mattsson L., 2011, MNRAS, 414, 781
\bibitem[\protect\citeauthoryear{Mattsson \& H\"ofner}{2011}]{Mattsson11b}
  Mattsson L. \& H\"ofner S., 2011, A\&A, 533, A42
\bibitem[\protect\citeauthoryear{Mattsson, Andersen \& Munkhammar}{2011}]{Mattsson11c}
  Mattsson L. Andersen A. C. \& Munkhammar J. D., 2011, MNRAS, submitted (Paper I)
\bibitem[\protect\citeauthoryear{McKee}{1989}]{McKee89}  
  McKee C. F., 1989, IAUS, 135, 431
\bibitem[\protect\citeauthoryear{Morgan \& Edmunds}{2003}]{Morgan03}
  Morgan, H.L. \& Edmunds, M.G. 2003, MNRAS, 343, 427
\bibitem[\protect\citeauthoryear{Morgan et al.}{2003}]{Morgan03b}
  Morgan H.L., Dunne L., Eales S.A., Ivison R.J., Edmunds M.G., 2003, ApJ, 597, L33
\bibitem[\protect\citeauthoryear{Moustakas et al.}{2010}]{Moustakas10}  
  Moustakas J., Kennicutt R. C., Tremonti C. A., et al. 2010, ApJS, 190, 233
\bibitem[\protect\citeauthoryear{Munoz-Mateos et al.}{2009}]{Munoz-Mateos09}  
  Mu\~noz-Mateos J. C., et al. 2009, ApJ, 701, 1965
\bibitem[\protect\citeauthoryear{Nozawa \& Kozasa}{2006}]{Nozawa06}   
  Nozawa T. \& Kozasa T., 2006, ApJ, 648, 435
\bibitem[\protect\citeauthoryear{Ormel et al.}{2099}]{Ormel09}   
  Ormel C. W. , Paszun D., Dominik C. \&  Tielens A. G. G. M., 2009, A\&A, 502, 845
\bibitem[\protect\citeauthoryear{Ossenkopf}{1993}]{Ossenkopf93}   
  Ossenkopf V., 1993, A\&A, 280, 617
\bibitem[\protect\citeauthoryear{Pagani et al.}{2010}]{Pagani10}   
  Pagani L., Steinacker J., Bacmann A., Stutz A. \& Henning T., 2010, Science, 329, 1622
\bibitem[\protect\citeauthoryear{Pagel}{1997}]{Pagel97} 
  Pagel B.E.J., 1997, "Nucleosynthesis and Chemical Evolution of Galaxies", Cambridge Univ. Press 
\bibitem[\protect\citeauthoryear{Pei}{1992}]{Pei92}    
  Pei Y. C., 1992, ApJ, 395, 130
\bibitem[\protect\citeauthoryear{Pilyugin \& Ferrini}{2000}]{Pilyugin00}
  Pilyugin L. S. \& Ferrini F., 2000, A\&A, 358, 72 
\bibitem[\protect\citeauthoryear{Pilyugin, V\'ilchez \& Contini}{2004}]{Pilyugin04}
  Pilyugin L. S., V\'ilchez J. M. \& Contini T., 2004, A\&A, 425, 849 
\bibitem[\protect\citeauthoryear{Pilyugin \& Thuan}{2005}]{Pilyugin05}
  Pilyugin L. S., \& Thuan T. X., 2005, ApJ, 631, 231
\bibitem[\protect\citeauthoryear{Pisano, Wilcots \& Elmegreen}{1998}]{Pisano98}
  Pisano D. J. , Wilcots E. M. \& Elmegreen B. G., 1998, ApJ, 115, 975
\bibitem [\protect\citeauthoryear{Rownd \& Young}{1999}]{Rownd99}  
  Rownd B. K., \& Young J. S., 1999, AJ, 118, 670
\bibitem [\protect\citeauthoryear{Schruba et al.}{2011}]{Schruba11}  
  Schruba A., Leroy A. K., Walter F., et al., 2011, ApJ, accepted, arxiv:1105.4605
\bibitem [\protect\citeauthoryear{Serra D\'iaz-Cano \& Jones}{2008}]{Serra08}    
  Serra D\'õaz-Cano L. \& Jones A. P., 2008, A\&A, 492, 127
\bibitem [\protect\citeauthoryear{Shen, Draine \& Johnson}{2008}]{Shen08}    
  Shen Y., Draine B. T. \& Johnson E. T., 2008, ApJ, 689, 260
\bibitem [\protect\citeauthoryear{Taylor \& Williams}{1993}]{Taylor93}  
  Taylor S. D. \& Williams D. A., 1993, MNRAS, 260, 280
\bibitem [\protect\citeauthoryear{Tielens et al.}{1987}]{Tielens87}  
  Tielens A. G. G. M., Seab C. G., Hollenbach D. J. \& McKee C. F., 1987, ApJ, 319, L109
\bibitem [\protect\citeauthoryear{Tielens et al.}{1994}]{Tielens94}  
  Tielens A. G. G. M., McKee C. F, Seab C. G., \& Hollenbach D. J., 1994, ApJ, 431, 321
\bibitem[\protect\citeauthoryear{Tricomi}{1947}]{Tricomi47}  
  Tricomi F. G., 1947, Annali di Matematica Pura ed Applicata: Serie Quarta, 26,141
\bibitem[\protect\citeauthoryear{Vollmer et al.}{2007}]{Vollmer07}   
  Vollmer C., Hoppe P., Brenker F. E., \& Holzapfel C., 2007, ApJ, 666, L49
\bibitem[\protect\citeauthoryear{Walter et al.}{2008}]{Walter08}    
  Walter F.,  Brinks E.,  de Blok W. J. G., Bigiel F., Kennicutt Jr., R.C., Thornley M.D. \& Leroy A., 2008, AJ, 136, 2563
\bibitem[\protect\citeauthoryear{Whittet}{1991}]{Whittet91} 
  Whittet D.C.B., 1991, Dust in the Galactic Environment, IOP Publishing, Bristol
\bibitem[\protect\citeauthoryear{Wong \& Blitz}{2002}]{Wong02}   
  Wong T., \& Blitz L. 2002, ApJ, 569, 157
\bibitem[\protect\citeauthoryear{Zhukovska, Gail \& Trieloff}{2008}]{Zhukovska08}   
  Zhukovska S., Gail H.-P. \& Trieloff M., 2008, A\&A, 479, 453  
\end{thebibliography}
\end{document}